\documentclass[10pt,twocolumn,twoside]{IEEEtran}
\ifCLASSINFOpdf
\fi
\usepackage{cite}
\usepackage{graphicx}
\usepackage{times}
\usepackage{latexsym}

\usepackage[mathscr]{eucal}
\usepackage[center]{caption2}
\usepackage{array}
\usepackage{fancyhdr}
\usepackage{amsmath,amssymb}
\usepackage{bm} 
\usepackage{subfigure}
\usepackage{citesort}
\usepackage{multirow}

\ifCLASSINFOpdf
\else
\fi

\usepackage{xcolor}

\makeatother

\usepackage{amsthm}
\usepackage{amssymb}
\usepackage{stfloats}
\usepackage{graphicx}
\usepackage{subfigure}
\usepackage{tabularx}
\usepackage{epsfig,epsf,color,balance,cite}
\usepackage{verbatim}
\usepackage{url}
\usepackage{bm}
\usepackage{subeqnarray}
\usepackage{cases}
\usepackage{tabularx}
\usepackage{booktabs}
\usepackage{algpseudocode}
\usepackage{algorithm}
\usepackage{amsfonts}

\begin{document}

%\title{Secret Key Generation Based on Intelligent Reflecting Surface In Multi-antenna Network}
\title{Intelligent Reflecting Surface-Assisted Secret Key Generation In Multi-antenna Network}

\author{You~Chen,
        Guyue~Li,~\IEEEmembership{Member,~IEEE},
        Cunhua~Pan,~\IEEEmembership{Member,~IEEE},
        Lei~Hu,~\IEEEmembership{Student Member,~IEEE},
        Aiqun~Hu,~\IEEEmembership{Member,~IEEE}

\thanks{Y. Chen, G. Li and L. Hu are with the School of Cyber Science and Engineering, Southeast University, Nanjing, China. (e-mail: {213170837, guyuelee, 220205311}@seu.edu.cn).}
\thanks{C. Pan is with the School of Electronic Engineering and Computer Science, Queen Mary University of London, London E1 4NS, U.K. (e-mail: c.pan@qmul.ac.uk).}
\thanks{A. Hu is with the National Mobile Communications Research Laboratory, Southeast University, Nanjing, 210096, China. (e-mail:aqhu@seu.edu.cn).}
\thanks{G. Li and A. Hu are also with the Purple Mountain Laboratories for Network and Communication Security, Nanjing, 210096, China.}
\thanks{G. Li is the corresponding author.}}
%\thanks{Y. Chen, G. Li and L. Hu are with the School of Cyber Science and Engineering, Southeast University, Nanjing, China. (e-mail: {213170837, guyuelee, 220205311}@seu.edu.cn). C. Pan is with the School of Electronic Engineering and Computer Science, Queen Mary University of London, London E1 4NS, U.K. (e-mail: c.pan@qmul.ac.uk). A. Hu is with the National Mobile Communications Research Laboratory, Southeast University, Nanjing, 210096, China. (e-mail:aqhu@seu.edu.cn). G. Li and A. Hu are also with the Purple Mountain Laboratories for Network and Communication Security, Nanjing, 210096, China. G. Li is the corresponding author.}}

%%%%%%%%%%%%%%%%%%%%%%%%%%%%%%%%%%%%%%%%%%%%%%%%%%%%%%%%%%%%%%%%%%%%%%%%%%%%%%%%%
%\markboth{Journal of \LaTeX\ Class Files,~Vol.~14, No.~8, April~2021}%



\maketitle
\begin{abstract}
Physical-layer key generation (PKG) can generate symmetric keys between two communication ends based on the reciprocal uplink and downlink channels. By smartly reconfiguring the radio signal propagation, intelligent reflecting surface (IRS) is able to improve the secret key rate of PKG. However, existing works involving IRS-assisted PKG are concentrated in single-antenna wireless networks. So this paper investigates the problem of PKG in the IRS-assisted multiple-input single-output (MISO) system, which aims to maximize the secret key rate by optimally designing the IRS passive beamforming. First, we analyze the correlation between channel state information (CSI) of eavesdropper and legitimate ends and derive the expression of the upper bound of secret key rate under passive eavesdropping attack. Then, an optimal algorithm for designing IRS reflecting coefficients based on Semi-Definite Relaxation (SDR) and Taylor expansion is proposed to maximize the secret key rate. Numerical results show that our optimal IRS-assisted PKG scheme can achieve much higher secret key rate when compared with two benchmark schemes.
\end{abstract}

\begin{IEEEkeywords}
Physical layer security, intelligent reflecting surface, secret key generation, fractional programming, semidefinite relaxation.
\end{IEEEkeywords}

\IEEEpeerreviewmaketitle

\section{Introduction}
\IEEEPARstart{P}{hysical-layer} key generation (PKG) is an enabling technology to improve the physical layer security in wireless communication networks \cite{9205612}. Based on the reciprocity of the uplink and downlink channels, PKG can generate symmetric keys between two communication ends for data encryption. However, the application of relays is needed in case that the direct link channel between legitimate communication ends is not good enough to generate keys, which may add to the transmit power consumption.

Recently, intelligent reflecting surface (IRS) is emerging as an energy efficient technique to tackle with this problem and reconfigure the radio signal propagation. IRS only reflects the received signals as a passive array, thus incurring no transmit power consumption\cite{9090356}. At the same time, by adaptively tuning the amplitudes and/or phase shifts of low-cost passive reflecting elements, IRS is capable of smartly altering the incident signals to add destructively with the non-IRS-reflected signal at the eavesdropper, which can decrease the information leakage to eavesdropper.
%energy efficient passive relay without ZF chains, has emerged as a promising technology to

In view of these characteristics, a few literatures have applied IRS in PKG\cite{9361290,9360860,9298937}. In \cite{9360860}, Z. Ji et al. proposed an one-time pad encrypted data transmission scheme based on random phase shifting of the IRS elements. In \cite{9361290}, X. Lu et al. effectively adjusted the switch state of IRS units to maximize the secret key rate in single-input single-putput (SISO) network. However, the correlation between eavesdropper's and the legitimate users' channel state information (CSI) was ignored. In \cite{9298937}, Z. Ji et al. considered channel correlation and optimally designed IRS reflecting coefficients to maximize the lower bound of secret key rate. To sum up, the existing works in IRS-assisted PKG have proved the performance advantages of utilizing IRS in PKG, but the problem of maximizing the upper bound of secret key rate under channel correlation has not been studied. Besides, existing BSs are usually equipped with multiple antennas to improve communication performance, while the field of IRS-assisted PKG in multi-antenna wireless communication networks still remains unexplored, where the complexity of solving the optimization algorithm will increase with the dimension of channel estimations.

In this paper, we will optimize IRS reflection coefficients in a multiple-input single-output single-antenna eavesdropper (MISOSE) network with the purpose of maximizing secret key rate. Our main contributions are as follows:
	\begin{itemize}
		\item We present an IRS-assisted MISOSE system model and derive the expression of the upper bound of secret key rate. The introduction of IRS brings in high correlation between the CSI of eavesdropper and that of legitimate communication ends, which can lead to an obvious decrease in secret key rate.
		\item We propose an algorithm for IRS coefficients to realize the maximal secret key rate. The optimization problem is non-convex and hence, we solve it by applying Taylor expansion and Semi-Definite Relaxation (SDR) techniques.
%		\item Simulation results demonstrate the influence of channel correlation on secret key rate and verify that the secret key rate of our proposed scheme is much higher than that of the other two benchmark schemes. With the growing number of IRS reflecting elements, we observe a significant increase in secret key rate. for the IRS reflecting coefficient vector
		\item Simulation results demonstrate the impact of channel correlation on secret key rate, and verify that our proposed scheme can outperform the other two benchmark schemes in improving both secret key rate and bit disagreement ratio (BDR).
% With the increasing number of IRS reflecting elements, we observe a significant increase in secret key rate.
	\end{itemize}

\section{system model and signal presentation}
%In this section, we first introduce our MISOSE system model and give the signal presentations of our IRS-assisted PKG scheme.
\subsection{System Model}
As shown in Fig. 1, we consider a general time-division duplex (TDD) MISOSE channel model with three nodes, where an $M$-antenna station (BS) generates secret keys with a single-antenna user terminal (UT) with the assistance of an IRS. The IRS is located between the BS and the UT, comprising $L$ reflecting elements arranged in a uniform Planar array (UPA) with $X$ rows and $Y$ columns. A single-antenna eavesdropper (Eve) intends to eavesdrop the secret keys based on his own channel observations and all the information exchanged over the public channel.
	\begin{figure}
		\centering
		\includegraphics[width=2.7in]{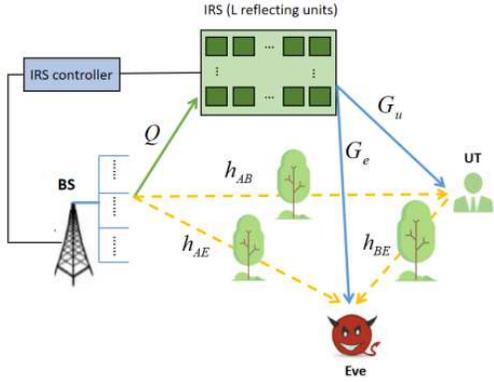}
		\caption{The MISOME communication system model.}
	\end{figure}
We consider a situation where the direct channels between three nodes are assumed to be partially blocked by obstacles. Since wireless channel has plenty of scatters, we model the small-scale fading of BS-to-UT, BS-to-Eve, and Eve-to-UT channels as Rayleigh fading\cite{han2019large}, namely $\mathbf h_{A B}$, $\mathbf h_{A E}$ and $\mathbf h_{B E}$. The elements of $\mathbf h_{i j},i,j \in \left\{A,B,E\right\}$ are independent and identically distributed (i.i.d.) in the complex Gaussian distribution with zero mean and unit variance. Considering the effect of path loss, the variance $\sigma_{i j}^{2}$ of corresponding channel equals to $\alpha_{i j}^{2}$, where $\alpha_{i j} \sim C \mathcal{N}\left(0,10^{-0.1 P L(d)}\right),i,j \in \left\{A,B,E\right\}$. The path loss can be obtained as $P L(d)[d B]=PL_{0}+10 c \log _{10}(d)$, $d$, $PL_{0}$ and $c$ denote the distance between transmitter and receiver, constant path loss term and path loss exponent respectively. With the assumption that the BS and IRS are located at a higher altitude, the BS-to-IRS channel can be modeled as a rank-one matrix \cite{rafieifar2020secrecy}
\begin{equation}
\mathbf Q=\sqrt{M L} \alpha_{Q} \mathbf a_{I R S}\left(\theta_{I R S}, \gamma_{I R S}\right) \mathbf a_{B S}^{H}\left(\varphi_{B S}\right)\label{Q},
\end{equation}
where $\alpha_{Q} \sim C \mathcal{N}\left(0,10^{-0.1 P L(d)}\right)$ denotes the complex gain of the BS-to-IRS channel, $\mathbf a_{I R S}\left(\theta_{I R S}, \gamma_{I R S}\right)$ is the normalized IRS array response vector and $\mathbf a_{B S}\left(\varphi_{B S}\right)$ is the normalized BS array response vector, that are denoted by
\begin{equation}
\begin{aligned}
\mathbf a_{I R S}\left(\theta_{I R S}, \gamma_{I R S}\right) =\frac{1}{\sqrt{L}}&\left[1,..., e^{\tau_{x,y}(\theta_{IRS},\gamma_{IRS})},...,\right.\\
&,...,\left.e^{\tau_{X,Y}(\theta_{I R S}, \gamma_{I R S})}\right]^{T}\label{array},
\end{aligned}
\end{equation}
\begin{equation}
\mathbf a_{B S}\left(\varphi_{B S}\right)=\frac{1}{\sqrt{M}}\left[1,...,e^{-j 2 \pi \frac{d}{\lambda}(M-1) \sin \varphi_{B S}}\right]^{T},
\end{equation}
where $\theta_{I R S}\in[0,\pi]$ and $\gamma_{I R S}\in[0,\pi]$ represent the elevation and azimuth angles of arrival (AOA) for the IRS and $\varphi_{B S} \in[0,2 \pi]$ is the azimuth angle of angle of departure (AOD) for the BS. $\tau_{x,y}(\theta,\gamma)=-j 2 \pi \frac{d}{\lambda}\left((x-1)\sin \theta \sin \gamma+(y-1)\cos \gamma\right)$ is a function of elevation AOA/AOD $\theta$ and azimuth AOA/AOD $\gamma$ of IRS. Similar to \eqref{Q}, the IRS-to-UT and IRS-to-Eve reflected channels can be respectively given by
\begin{small}
\begin{equation}
\mathbf G_{U}=\sqrt{L} \alpha_{G_{U}} \mathbf a_{I R S, U}^{H}\left(\varphi_{I R S, U}, \omega_{I R S, U}\right),
\end{equation}
\end{small}
\begin{small}
\begin{equation}
\mathbf G_{E}=\sqrt{L} \alpha_{G_{E}} \mathbf a_{I R S, E}^{H}\left(\varphi_{I R S, E}, \omega_{I R S, E}\right),
\end{equation}
\end{small}
where $\alpha_{G_{U}} \sim C \mathcal{N}\left(0,10^{-0.1 P L(d)}\right)$ and $\alpha_{G_{E}} \sim C \mathcal{N}\left(0,10^{-0.1 P L(d)}\right)$ are the complex gain of the IRS-to-UT and IRS-to-Eve channel respectively, $\mathbf a_{I R S, U}\left(\varphi_{I R S, U}, \omega_{I R S, U}\right)$ and $\mathbf a_{I R S, E}\left(\varphi_{I R S, E}, \omega_{I R S, E}\right)$ are the normalized array response vectors of the IRS associated with the IRS-to-UT and the IRS-to-Eve paths respectively, definitions of them are similar to \eqref{array}. $\varphi_{I R S, U}\in[0,\pi]$ and $\omega_{I R S, U} \in[0,\pi]$ represent the elevation and azimuth AOD from IRS to UT, $\varphi_{I R S, E} \in[0,\pi]$ and $\omega_{I R S, E} \in[0,\pi]$ represent the elevation and azimuth AOD from IRS to Eve.
%\begin{equation}
%\begin{array}{c}
%a_{I R S, U}\left(\varphi_{I R S, U}, \omega_{I R S, U}\right)=\frac{1}{\sqrt{L}}\left[1, \ldots, e^{-j 2 \pi \frac{d}{\lambda}\left(x \sin \varphi_{I R S, U} \sin \omega_{I R S, U}+y \cos \omega_{I R S, U}\right)}, \ldots,\right. \\
%\left., \ldots, e^{-j 2 \pi \frac{d}{\lambda}\left((\sqrt{L}-1) \sin \varphi_{I R S, U} \sin \omega_{I R S, U}+(\sqrt{L}-1) \cos \omega_{I R S, U}\right)}\right]^{T}
%\end{array}
%\end{equation}
%\begin{equation}
%\begin{array}{c}
%a_{I R S, E}\left(\varphi_{I R S, E}, \omega_{I R S, U}\right)=\frac{1}{\sqrt{L}}\left[1, \ldots, e^{-j 2 \pi \frac{d}{\lambda}\left(x \sin \varphi_{I R S, E} \sin \omega_{I R S, E}+y \cos \omega_{I R S, E}\right)}, \ldots,\right. \\
%\left., \ldots, e^{-j 2 \pi \frac{d}{\lambda}\left((\sqrt{L}-1) \sin \varphi_{I R S, E} \sin \omega_{I R S, E}+(\sqrt{L}-1) \cos \omega_{I R S, E}\right)}\right]^{T}
%\end{array}
%\end{equation}
Referring to \cite{9298937}, we define the reflecting coefficient vector of IRS as $\mathbf v=\left[v_{1}, \ldots, v_{\mathrm{L}}\right]^{T}$ . The $n$th element of $\mathbf v$ is $v_{n}=e^{j \phi_{n}}$, where $\phi_{n} \in[0,2 \pi]$ is the phase shifts on the incident signal by IRS's $n$th element, $n=1,...,L$.

Considering that BS is always located at a higher altitude and far away from Eve, $\mathbf G_{U}$ and $\mathbf Q$ are independent of each other, contributing to the the independence between uplink channel estimation of Eve and legitimate communication ends. Besides, the uplink channel estimation at Eve side is a scalar and poses little threats of information leakage. Based on the above analysis, we only consider the correlation between UT's and Eve's downlink channel estimations in this paper.
\subsection{Framework of IRS-assisted PKG Scheme}
In this part, we will give signal presentations of our IRS-assisted PKG scheme. The PKG process generally contains four stages, namely channel probing, quantization, information reconciliation, and privacy amplification \cite{9205612}, this paper mainly focuses on the first step. During the channel probing step, BS and UT will alternately exchange pilots and perform channel estimations in TDD mode. Moreover, as the uplink and downlink channels are reciprocal, the uplink channel matrix can be represented by the transpose of downlink channel matrix.

Firstly, in the downlink channel probing, the BS broadcasts the orthogonal pilot signal $\mathbf S^{D L} \in C^{M \times T_{D}}$, where $T_{D}$ is the length of pilot and $\mathbf S^{D L}(\mathbf S^{D L})^{H} = \mathbf I$, $P$ is the transmit power of the BS. So the received signals of UT and Eve are given by
\begin{equation}
\mathbf Y_{U}^{D L}=\sqrt{P}\left(\mathbf h_{A B}+\mathbf v^{T} \mathbf R_{U}\right) \mathbf S^{D L}+\mathbf N_{U}^{D L},
\end{equation}
\begin{equation}
\mathbf Y_{E}^{D L}=\sqrt{P}\left(\mathbf h_{A E}+\mathbf v^{T} \mathbf R_{E}\right) \mathbf S^{D L}+\mathbf N_{E}^{D L},
\end{equation}where $\mathbf R_{U}=\operatorname{diag}(\mathbf G_{U}) \mathbf Q$ and $\mathbf R_{E}=\operatorname{diag}(\mathbf G_{E}) \mathbf Q$. $\mathbf N_{U}^{D L}$ and $\mathbf N_{E}^{D L}$ are complex additive white Gaussian noise (AWGN) with variance $\sigma_{1}^{2}$ and $\sigma_{2}^{2}$ of each elements. By the least square (LS) estimation, UT and Eve estimate their respective downlink CSI as
\begin{equation}
\mathbf Z_{U}^{D L}=\frac{1}{\sqrt{P}}\mathbf Y_{U}^{D L}\left(\mathbf S^{D L}\right)^{\mathrm{H}}=\mathbf h_{A B}+\mathbf v^{T} \mathbf R_{U}+\mathbf N_{U}^{D L'}\,
\end{equation}
\begin{equation}
\mathbf Z_{E}^{D L}=\frac{1}{\sqrt{P}}\mathbf Y_{E}^{D L}\left(\mathbf S^{D L}\right)^{\mathrm{H}}=\mathbf h_{A E}+\mathbf v^{T} \mathbf R_{E}+\mathbf N_{E}^{D L'},
\end{equation}where the variance of each elements in AWGN $\mathbf N_{U}^{D L'}$ and $\mathbf N_{E}^{D L'}$ are $\sigma_{1}^{2}/P$ and $\sigma_{2}^{2}/P$.

Secondly, in the uplink channel probing, the UT broadcasts the orthogonal pilot signal $\mathbf S^{U L} \in C^{M \times T_{U}}$, where $T_{U}$ is the length of pilot and $\mathbf S^{U L}(\mathbf S^{U L})^{H} = \mathbf I$, the transmit power of the BS is also $P$. So the received signal of BS is given by
\begin{equation}
\mathbf Y_{U}^{U L}=\sqrt{P}\left(\mathbf h_{A B}^{T}+\mathbf R_{U}^{T} \mathbf v\right) \mathbf S^{U L}+\mathbf N_{U}^{U L},
\end{equation}where $\mathbf N_{U}^{U L}$ is complex AWGN with variance $\sigma_{1}^{2}$ of each elements due to channel reciprocity. By the LS estimation, BS estimate its respective downlink CSI as
\begin{equation}
\mathbf Z_{U}^{U L}=\frac{1}{\sqrt{P}}\mathbf Y_{U}^{U L}\left(\mathbf S^{U L}\right)^{\mathrm{H}}=\mathbf h_{A B}^{T}+\mathbf R_{U}^{T} \mathbf v+\mathbf N_{U}^{U L'},
\end{equation}
and the variance of each elements in AWGN $\mathbf N_{U}^{U L'}$ is $\sigma_{1}^{2}/P$.

Finally, BS and UT vectorize the estimated effective channel matrices $\mathbf H_{A}=\operatorname{vec}\left(\mathbf Z_{U}^{D L}\right)=\mathbf h_{A B}^{T}+\mathbf R_{U}^{T} \mathbf v+\left(\mathbf N_{U}^{D L'}\right)^{T}$ and $\mathbf H_{B}=\operatorname{vec}\left(\mathbf Z_{U}^{U L}\right)=\mathbf h_{A B}^{T}+\mathbf R_{U}^{T} \mathbf v+\mathbf N_{U}^{U L'}$
to generate the secret key. Similarly, Eve can also obtain the vectorized channel estimation $\mathbf H_{E}=\operatorname{vec}\left(\mathbf Z_{E}^{D L}\right)=\mathbf h_{A E}^{T}+\mathbf R_{E}^{T} \mathbf v+\left(\mathbf N_{E}^{D L'}\right)^{T}$.

Based on the above analytical expressions, we observed that there exists correlation between Eve's and legitimate ends' vectorized channel estimations due to the common BS-to-IRS channel $\mathbf Q$ and IRS reflecting coefficient vector $\mathbf v$. This will increase information leakage to Eve, thus we must optimally design $\mathbf v$ to maximize the secret key rate.
\section{secret key rate analysis}
In this section, we first derive the expression of the upper bound of secret key rate in our IRS-assisted wireless communication network, then discuss the impact of channel correlation on secret key rate.
%\subsection{The Secret Key Rate In Presence of Eve}
%We note that in non-IRS systems with potential eavesdropping from Eve, when Eve is located half a wavelength away or more from BS and UT, Eve's CSI  can be treated as independent to the CSI of BS and UT and thus neglected in the secret key rate characterization \cite{li2018high}. However, according to our analy, due to the existence of common channel BS-to-IRS $Q$ and , Eve and legitimate communication ends (BS and UT) will observe correlated CSI in our IRS-assisted key generation system.
\subsection{Formulation of Secret Key Rate}
We assume that the statistical CSI of all channels are known at BS\cite{9298937}. By invoking the central limit theorem, the composite channels can be approximated by the Gaussian distribution for a sufficiently large number of $L$ \cite{9360860}. Based on our model in Section \uppercase\expandafter{\romannumeral2}, the upper bound of secret key rate of legitimate CSI $\mathbf H_{A}$ and $\mathbf H_{B}$ under eavesdropping channel $\mathbf H_{E}$ can be expressed as \cite{1993}
\begin{small}
\begin{equation}
\begin{aligned}
R(\mathbf v)&=I\left(\mathbf H_{A} ; \mathbf H_{B}\mid \mathbf H_{E}\right) \\
&=\log _{2} \frac{\operatorname{det}\left(\bm{\Re}_{A E}\right) \operatorname{det}\left(\bm{\Re}_{B E}\right)}{\operatorname{det}\left(\bm{\Re}_{E}\right) \operatorname{det}\left(\bm{\Re}_{A B E}\right)}\label{12},
\end{aligned}
\end{equation}
\end{small}where $\mathbf \Re_{x...y}=E\left\{\left[\begin{array}{llll}\mathbf H_{x}&...&\mathbf H_{y}\end{array}\right]^{T}\left[\begin{array}{llll}\mathbf H_{x}^{H}&...& \mathbf H_{y}^{H}\end{array}\right]\right\}$ and $\mathbf K_{x y}=E\{\mathbf H_{x} \mathbf H_{y}^{H}\}$, $x,...,y \in \left\{A,B,E\right\}$ are correlation matrices, $\operatorname{det}\{.\}$ is the determinant of matrix.
%$\Re_{x y \ldots z}=E\left\{\left[\begin{array}{llll}H_{x} & H_{y} & \ldots & H_{z}\end{array}\right]^{T}\left[\begin{array}{llll}H_{x}^{H} & H_{y}^{H} & \ldots & H_{z}^{H}\end{array}\right]\right\}$ and $K_{x y}=E\{H_{x} H_{y}^{H}\}$,$x,y,...,z \in \left\{A,B,E\right\}$correlation matrix

Furthermore, we define correlation matrices of IRS reflected channels as $\mathbf W_{U}=E\{\mathbf R_{U}^{T} \mathbf v \mathbf v^{H} \mathbf R_{U}^{*}\}=p_{U}(\mathbf v) \mathbf R_{BS}$,$\mathbf W_{E}=E\{\mathbf R_{E}^{T} \mathbf v \mathbf v^{H} \mathbf R_{E}^{*}\}=p_{E}(\mathbf v) \mathbf R_{BS}$ and $\mathbf W_{L}=E\{\mathbf R_{U}^{T} \mathbf v \mathbf v^{H} \mathbf R_{E}^{*}\}=p_{L}(\mathbf v) \mathbf R_{BS}$, where
scalar $p_{U}(\mathbf v)=\left|\alpha_{G_{U}} \alpha_{Q}\right|^{2} \mathbf v^{T} \bm{\beta} \bm{\beta}^{\mathrm{H}} \mathbf v^{*}$, $p_{E}(\mathbf v)=\left|\alpha_{G_{E}} \alpha_{Q}\right|^{2} \mathbf v^{T} \bm{\psi} \bm{\psi}^{\mathrm{H}} \mathbf v^{*}$, $p_{L}(\mathbf v)=\alpha_{G_{U}} \alpha_{G_{E}}^{*} \left|\alpha_{Q}\right|^{2} \mathbf v^{T} \bm{\beta} \bm{\psi}^{\mathrm{H}} \mathbf v^{*}$ and
\begin{small}
\begin{equation}
\label{R_BS}
\begin{aligned}
\mathbf R_{B S}=& \mathrm{E}\left\{\left[1, e^{j 2 \pi \frac{d}{2} \sin \varphi_{B S}}, \ldots, e^{j 2 \pi \frac{d}{\lambda}(M-1) \sin \varphi_{B S}}\right]^{T}\right.\\
&\left.\bullet\left[1, e^{-j 2 \pi \frac{d}{2} \sin \varphi_{B S}}, \ldots, e^{-j 2 \pi \frac{d}{2}(M-1) \sin \varphi_{B S}}\right]\right\}.
\end{aligned}
\end{equation}
\end{small}Vectors $\bm{\beta}=\left[e^{\beta_{1,1}}, \ldots, e^{\beta_{X,Y} }\right]^{T}$ and $\bm{\psi}=\left[e^{\psi_{1,1}}, \ldots, e^{\psi_{X,Y} }\right]^{T}$ are related to the positions of UT and Eve respectively,
\begin{equation}
\begin{aligned}
\beta_{x, y}=\tau_{x,y}(\theta_{IRS},\gamma_{IRS})-\tau_{x,y}(\varphi_{I R S, U}, \omega_{I R S, U}),
\end{aligned}
\end{equation}
\begin{equation}
\begin{aligned}
\psi_{x, y}=\tau_{x,y}(\theta_{IRS},\gamma_{IRS})-\tau_{x,y}(\varphi_{I R S, E}, \omega_{I R S, E}).
\end{aligned}
\end{equation}
Thus, it can be derived that $\mathbf K_{A A}=\mathbf K_{B B}=\mathbf W_{U}+\sigma_{U}^{2}\mathbf I$, $\mathbf K_{E E}=\mathbf W_{E}+\sigma_{E}^{2}\mathbf I$, $\mathbf K_{A B}=\mathbf K_{B A}=\mathbf W_{U}+\sigma_{A B}^{2}I$ and $\mathbf K_{A E}=\mathbf K_{B E}=\mathbf K_{E A}^{H}=\mathbf K_{E B}^{H}=\mathbf W_{L}$, where $\sigma_{U}^{2}=\sigma_{1}^{2}/P+\sigma_{A B}^{2}$ , $\sigma_{E}^{2}=\sigma_{2}^{2}/P+\sigma_{A E}^{2}$ and $\sigma_{N}^{2}=\sigma_{1}^{2}/P+2\sigma_{A B}^{2}$. Thus, the determinant of matrix terms in \eqref{12} is shown as
\begin{equation}
\begin{aligned}
\operatorname{det}(&\left.\bm{\Re}_{E}\right)= \operatorname{det}\left(\mathbf W_{E}+\sigma_{E}^{2} \mathbf I\right)\label{22}.
\end{aligned}
\end{equation}
\begin{equation}
\begin{aligned}
\operatorname{det}(&\left.\bm{\Re}_{A B E}\right)=\left(\sigma_{1}^{2}\right)^{M} \operatorname{det}\left(\mathbf W_{E}+\sigma_{E}^{2} \mathbf I\right) \\
& \bullet \operatorname{det}\left(2 \mathbf W_{U}+\sigma_{N}^{2} \mathbf I-2 \mathbf W_{L}\left(\mathbf W_{E}+\sigma_{E}^{2} \mathbf I\right)^{-1} \mathbf W_{L}^{H}\right)\label{20},
\end{aligned}
\end{equation}
\begin{equation}
\begin{aligned}
\operatorname{det}\left(\bm{\Re}_{A E}\right) &=\operatorname{det}\left(\bm{\Re}_{B E}\right)=\operatorname{det}\left(\mathbf W_{U}+\sigma_{U}^{2} \mathbf I\right) \\
\bullet & \operatorname{det}\left(\mathbf W_{E}+\sigma_{E}^{2} \mathbf I-\mathbf W_{L}^{H}\left(\mathbf W_{U}+\sigma_{U}^{2} \mathbf I\right)^{-1} \mathbf W_{L}\right)\label{21},
\end{aligned}
\end{equation}
It is worth noting that the rank of $\mathbf W_{U}$, $\mathbf W_{E}$ and $\mathbf W_{L}$ are all equal to $1$, which means that they all have at most one non-zero eigenvalue. Therefore, we can utilize the properties that the determinant of a given matrix can be expressed as the product of all eigenvalues of the matrix to transform $R$ into the following form.

	\emph{Theorem 1:}
	By substituting the parameters of channel model into expression \eqref{12}, the secret key rate can be calculated as
\begin{small}
\begin{equation}
\begin{aligned}
\label{theorem1}
R(\mathbf v)=& \log _{2} \frac{\left(\sigma_{E}^{2} M p_{U}(\mathbf v)+\sigma_{U}^{2} M p_{E}(\mathbf v)+\sigma_{E}^{2} \sigma_{U}^{2}\right)^{2}}{\left(M p_{E}(\mathbf v)+\sigma_{E}^{2}\right)\left(2 \sigma_{E}^{2} M p_{U}(\mathbf v)+\sigma_{N}^{2} M p_{E}(\mathbf v)+\sigma_{E}^{2} \sigma_{N}^{2}\right)} \\
&+\log _{2} \frac{\left(\sigma_{U}^{2}\right)^{2 M-2}}{\left(\sigma_{1}^{2}\right)^{M}\left(\sigma_{N}^{2}\right)^{M-1}} \\
=& \log _{2} \frac{f(\mathbf v)}{g(\mathbf v)}+\log _{2} \frac{\left(\sigma_{U}^{2}\right)^{2 M-2}}{\left(\sigma_{1}^{2}\right)^{M}\left(\sigma_{N}^{2}\right)^{M-1}}.
\end{aligned}
\end{equation}
\end{small}According to the expressions of $p_{U}(\mathbf v)$, $p_{E}(\mathbf v)$ and $p_{L}(\mathbf v)$, the numerator $f(\mathbf v)$ and denominator $g(\mathbf v)$ in \eqref{theorem1} are all functions of $\mathbf v$,
\begin{equation}
\begin{aligned}
&f(\mathbf v)= \sigma_{U}^{4} \sigma_{E}^{4}+2 M^{2} \sigma_{E}^{2} \sigma_{U}^{2} p_{U}(\mathbf v) p_{E}(\mathbf v)+M^{2} \sigma_{E}^{4} p_{U}(\mathbf v)^{2}\\
&+2 M \sigma_{U}^{4} \sigma_{E}^{2} p_{E}(\mathbf v)+M^{2} \sigma_{U}^{4} p_{E}(\mathbf v)^{2}+2 M \sigma_{U}^{2} \sigma_{E}^{4} p_{U}(\mathbf v),
\end{aligned}
\end{equation}
\begin{equation}
\begin{aligned}
&g(\mathbf v)= \sigma_{N}^{4} \sigma_{E}^{4}+M^{2} \sigma_{N}^{2} p_{E}(\mathbf v)^{2}+2 M^{2} \sigma_{E}^{2} p_{U}(\mathbf v) p_{E}(\mathbf v) \\
&+2 M \sigma_{E}^{4} p_{U}(\mathbf v)+2 M \sigma_{N}^{2} \sigma_{E}^{2} p_{E}(\mathbf v).
\end{aligned}
\end{equation}
	\emph{Proof:}
	See Appendix A. $\hfill \blacksquare$
\subsection{Discussion}
In the PKG scheme without IRS, the upper bound of secret key rate $R=2M \log _{2} \left(\sigma_{U}^{2}\right)-M \log _{2}\left(\sigma_{1}^{2}\sigma_{N}^{2}\right)$ is independent of $\mathbf v$. While in our IRS-assisted system, IRS reflecting coefficient vector $\mathbf v$ can be well designed to increase the value of fractional term $f(\mathbf v)/g(\mathbf v)$ in \eqref{theorem1}, thus improving the secret key rate. However, due to the existence of channel correlation brought by IRS, the secret key rate is reduced from $R_{\rm nocorr}(\mathbf v)=I\left(\mathbf H_{A} ; \mathbf H_{B}\right)$ to $R(\mathbf v)=I\left(\mathbf H_{A} ; \mathbf H_{B}\mid \mathbf H_{E}\right)$. Therefore, we define the reduction of maximal secret key rate as
\begin{equation}
R_{r}=\frac{\left[R(\mathbf v)\right]_{\rm max}-\left[R_{\rm nocorr}(\mathbf v)\right]_{\rm max}}{\left[R_{\rm nocorr}(\mathbf v)\right]_{\rm max}}\times 100\%,
\end{equation}
where $\left[.\right]_{\rm max}$ denotes the maximal value of a function. $R_{r}$ represents the negative impact of correlation on secret key rate, the relationship between $R_{r}$ and parameters $L$ and $M$ will be demonstrated by simulation results in Section \uppercase\expandafter{\romannumeral5}.
%, since the optimal solution $\mathbf v_{corr}$ to $R(v)$ may not equal to $\mathbf v_{nocorr}$.
\section{optimization problem and solution}
In this section, we first present our optimization problem and then propose an algorithm to design the IRS reflecting coefficients that maximizes the secret key rate.
%In this section, we will present our optimization problem and propose an algorithm to design the optimal IRS reflecting coefficient vector $\mathbf v$ that maximizes the secret key rate.
\subsection{Optimization Problem}
According to \eqref{theorem1}, since $\operatorname{log}_{2}(.)$ is a monotonically increasing function, the optimization problem can be formulated as
\begin{small}
\begin{subequations}
\label{problem}
\begin{align}
&\max\limits_{v} \frac{f(\mathbf v)}{g(\mathbf v)} \label{problema}\\
&\text {s.t.} \, \mathbf v^{H} \mathbf E_{n} \mathbf v=1, n=1, \ldots, L \label{problemb},
\end{align}
\end{subequations}
\end{small}
where $\mathbf E_{n}=\mathbf e_{n} \mathbf e_{n}^{H}$ is a base vector whose $n$th element is 1 and others 0, so that $\left|v_{n}\right|^{2}=\left|\mathbf v^{\mathrm{H}} \mathbf e_{n}\right|^{2}=\mathbf v^{\mathrm{H}} \mathbf e_{n} \mathbf e_{n}^{\mathrm{H}} \mathbf v=\mathbf v^{\mathrm{H}} \mathbf E_{n} \mathbf v=1$ [4]. It is difficult to find the optimal solution of the problem in \eqref{problem}, because \eqref{problema} and \eqref{problemb} are both non-convex with respect to the optimization variable $\mathbf v$. Therefore, we combine the techniques of fractional programming, SDR method and Taylor expansion to transform the initial problem in \eqref{problem} into a convex problem.
%, and we further construct an approximate optimal solution to the initial problem by utilizing gaussian randomization.
\subsection{Solution to the Optimization Problem}
The detailed steps of our algorithm are listed as follows.

\emph{Step1:} First, we define $\mathbf V=(\mathbf v^{T})^{H}\mathbf v^{T}$, $\mathbf B=\bm{\beta}\bm{\beta}^{H}$ and $\mathbf X=\bm{\psi}\bm{\psi}^{H}$, $\mathbf V$, $\mathbf B$ and $\mathbf X$ are all rank-1 positive semidefinite matrices. Then by using $\mathbf v^{H}\mathbf A\mathbf v\mathbf v^{H}\mathbf B \mathbf v=\operatorname{Tr}(\mathbf A\mathbf V\mathbf B\mathbf V)$, $\mathbf v^{H}\mathbf A\mathbf v=\operatorname{Tr}(\mathbf A\mathbf V)$ where $\mathbf A$, $\mathbf B$ and $\mathbf V$ are any positive semi-definite matrices and $\operatorname{Tr}(.)$ is the trace of the matrix, we can transform $f(\mathbf v)$ and $g(\mathbf v)$ into convex function expressions $f(\mathbf V)$ and $g(\mathbf V)$ w.r.t. optimization variable $\mathbf V$.

\emph{Step2:} We introduce an auxiliary variable $\mu^{(t+1)}=\frac{f\left(\mathbf V^{(t)}\right)}{g\left(\mathbf V^{(t)}\right)}$ to transform the original fractional objective function into form of the Difference of Convex (DC) functions, where $t$ is the number of iterations\cite{shen2018fractional}. The problem in \eqref{problem} is transformed into the following equivalent non-fractional form
\begin{small}
\begin{subequations}
\label{newproblem}
\begin{align}
\max \limits_{\mathbf V} & f(\mathbf V)-\mu g(\mathbf V) \label{newproblema}\\
\text { s.t. } & \mathbf V \succeq0, \label{newproblemb}\\
& \operatorname{Tr}\left(\mathbf E_{n} \mathbf V\right)=1, n=1, \ldots, L \label{newproblemc}\\
& \operatorname{rank}(\mathbf V)=1 \label{newproblemd}.
\end{align}
\end{subequations}
\end{small}
\emph{Step3:} Since $f(\mathbf V)$ and $g(\mathbf V)$ are both convex functions w.r.t. optimization variable $\mathbf V$, the objective function \eqref{newproblema} is in form of DC functions. In order to further overcome the non-convexity of \eqref{newproblema}, we apply the Taylor expansion at $\mathbf V^{(m)}$,
\begin{small}
\begin{equation}
f(\mathbf V)=f\left(\mathbf V^{(m)}\right)+\operatorname{Tr}\left(\operatorname{Re}\left\{\nabla f\left(\mathbf V^{(m)}\right)^{H}\left(\mathbf V-\mathbf V^{(m)}\right)\right\}\right).
\end{equation}
\end{small}After that, $f(\mathbf V)-\mu g(\mathbf V)$ is turned into a concave function so that the convex optimization problem in \eqref{newproblem} can be solved using CVX toolbox. In view of the non-convex constraint $rank(\mathbf V)=1$,
we can drop the constraint first \cite{lyu2020irs}.

\emph{Step4:} However, the solution obtained from problem in \eqref{newproblem} solved by the CVX is generally not a rank-one solution. Thus, Gaussian randomization method can be utilized to construct an approximate solution for problem in \eqref{problem} based on the solution from problem in \eqref{newproblem} \cite{random}.

\section{simulation results}
%In this section, we presents simulation results to demonstrate the performance of our proposed IRS-assisted PKG scheme in a MISOSE network. BS, UT, Eve and the central point of IRS are located at (5, 0, 20), (0, 100, 0), (0,105,0) and (0, 100, 20). The constant path loss term is $PL_{0} = 30 dB$ at reference distance $d_{0} = 1 m$. The path loss exponents for the BS-to-UT (BS-to-Eve) channel, BS-to-IRS channel and IRS-to-UT (IRS-to-Eve) channel are $c_{AB} = c_{AE} = 3.5$, $c_{Q} = 2$ and $c_{G_{U}}= c_{G_{E}}=2$, respectively. Other parameters are set as: the carrier frequency $f = 1 GHz$, $P=20dBm$, $\sigma_{1}^{2}=\sigma_{2}^{2}=-80dBm$, $\frac{d}{\lambda}=0.1$, $M=4$ and $L=20$ if not specified otherwise. We compare the performance of our proposed PKG scheme with the other two benchmark schemes: (1) PKG without IRS; (2) PKG with with random IRS phase shifting.
In this section, we present simulation results to demonstrate the performance of our proposed IRS-assisted PKG scheme in a MISOSE network. BS, UT, Eve and the central point of IRS are located at (5, 0, 20), (0, 100, 0), (0,105,0) and (0, 100, 20). The constant path loss term is $PL_{0} = 30 dB$ at reference distance $d_{0} = 1 m$ and the path loss exponents are $c_{AB} = c_{AE} = 3.5$, $c_{Q} = 2$ and $c_{G_{U}}= c_{G_{E}}=2$. Other parameters are set as: $P=20dBm$, $\sigma_{1}^{2}=\sigma_{2}^{2}=-80dBm$, $\frac{d}{\lambda}=0.1$, $M=4$ and $L=20$ if not specified otherwise. We compare the performance of our proposed PKG scheme with the other two benchmark schemes: (1) PKG without IRS; (2) PKG with random IRS phase shifting.

	\begin{figure}
		\centering
		\includegraphics[width=2.7in]{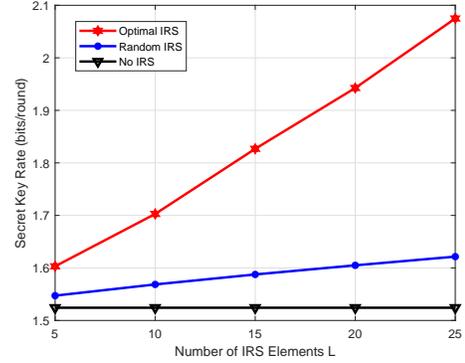}
		\caption{The secret key rate versus number of IRS elements $L$.}
	\end{figure}

Fig. 2 compares the secret key rate of the three key generation schemes versus the number of IRS reflecting elements $L$. Since the introduction of IRS can bring a new degree of freedom into secret key generation, the secret key rate of two IRS-assisted schemes has been greatly improved compared with that of the PKG scheme without IRS. At the same time, it is observed that our proposed scheme with optimal IRS phase shifting remarkably outperforms the benchmark scheme with random IRS phase shifting in improving key rate. This is because that the benchmark scheme lacks statistical CSI to optimize the passive beamforming at IRS, so that the application of IRS with random phase shifting can not bring maximal performance improvement. Besides, it is worth noting that the increase of IRS elements number $L$ can bring about an obvious improvement in secret key rate.

The negative impact of channel correlation is shown in Fig. 3, the channel correlation between Eve and legitimate ends increases with the increasing number of IRS reflecting elements and BS antennas, thus leading to more information leakage to Eve and a greater reduction of secret key rate $R_{r}$. Therefore, it is necessary to optimize the reflecting coefficients of IRS based on full consideration of channel correlation.
	\begin{figure}
		\centering
		\includegraphics[width=2.7in]{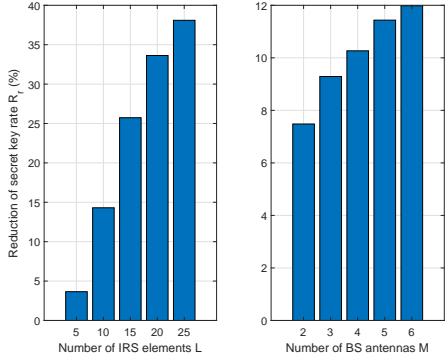}
		\caption{Secret key rate reduction $R_{r}$ versus $L$ and $M$.}
	\end{figure}

After channel probing, we normalize the channel observations and apply 4-bit CQG quantization algorithm with guardband $\delta=0.1$ and $0.2$ to generate initial keys. Fig. 4 compares the BDR versus transmit power $P$ under three PKG schemes, where BDR represents the ratio of the number of disagreement bits to the number of total bits of the initial secret keys. The introduction of IRS reflected channels increases the total channel gains, thus decreasing the BDR of our proposed scheme of optimal IRS phase shifting, which means that BS and UT can get a pair of initial keys with better quality.
	\begin{figure}
		\centering
		\includegraphics[width=2.7in]{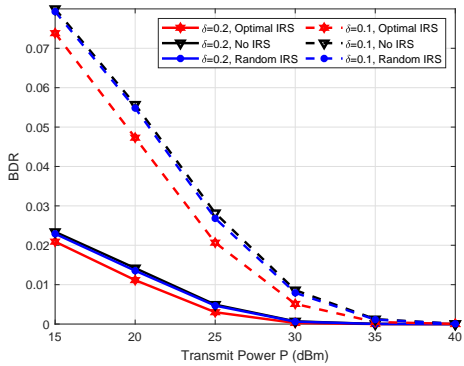}
		\caption{The BDR versus transmit power $P$ of three PKG schemes.}
	\end{figure}
\section{conclusion}
This paper investigated the problem of secret key rate optimization in the IRS-assisted MISOSE key generation system. First, the framework of IRS-assisted PKG scheme was presented. Then, we analyzed the channel correlation of eavesdropper and legitimate ends and derived the expression of the upper bound of secret key rate. After that, an optimization algorithm based on SDR and Taylor expansion was proposed to maximize the key rate by optimizing the IRS reflecting coefficients. Numerical results demonstrated the impact of channel correlation on secret key rate and verified that our proposed PKG scheme outperforms the other benchmark schemes.
%This paper investigated the problem of secret key rate optimization in the IRS-assisted MISO key generation system. We first analyzed the channel correlation of eavesdropper and legitimate ends and derived the expression of the upper bound of secret key rate under passive eavesdropping attack. Based on our analytical expressions, an optimization algorithm based on SDR and Taylor expansion was proposed to maximize the key rate by optimizing the IRS reflecting coefficients. Numerical results demonstrated the influence of channel correlation on secret key rate and verified that our proposed IRS-assisted PKG scheme outperformed the other benchmark schemes.

% %%%%%%%%%%%%%%%%%%%%%%%%%if have a single appendix:%%%%%%%%%%%%%%%%%%%%%%%%%%%
% or


\appendices
\section{Proof of Theorem 1}
%\begin{equation}
%\begin{array}{l}
%R=\log _{2} \frac{\operatorname{det}\left(\mathbf W_{U}+\sigma_{U}^{2} \mathbf I\right)^{2} \operatorname{det}\left(\mathbf W_{E}+\sigma_{E}^{2} \mathbf I-\mathbf W_{L}^{H}\left(\mathbf W_{U}+\sigma_{U}^{2} \mathbf I\right)^{-1} \mathbf W_{L}\right)^{2}}{\operatorname{det}\left(\mathbf W_{E}+\sigma_{E}^{2} \mathbf I\right)^{2} \operatorname{det}\left(2 \mathbf W_{U}+\sigma_{U}^{2} \mathbf I-2 \mathbf W_{L}\left(\mathbf W_{E}+\sigma_{E}^{2} \mathbf I\right)^{-1} \mathbf W_{L}^{H}\right)} \\
%-M \log _{2}\left(\sigma_{U}^{2}\right).
%\end{array}
%\end{equation}
%Here, $\mathbf W_{U}$, $\mathbf W_{E}$ and $\mathbf W_{L}$ are matrices with dimension of $M \times M$, which adds to the complexity of maximizing $R$.
After substituting \eqref{20}-\eqref{22} into \eqref{12}, we then utilize the properties that the determinant of a given matrix can be expressed as the product of all eigenvalues of the matrix to further transform \eqref{12} into the following form
\begin{small}
\begin{equation}
\label{R}
\begin{aligned}
\begin{aligned}
R &=2 \sum_{i=1}^{M} \log _{2} \lambda_{i}\left(\mathbf W_{U}+\sigma_{U}^{2} \mathbf I\right)-2 \sum_{i=1}^{M} \log _{2}\lambda_{i}\left(\mathbf W_{E}+\sigma_{E}^{2} \mathbf I\right)\\
&+2 \sum_{i=1}^{M} \log _{2} \lambda_{i}\left(\mathbf W_{E}+\sigma_{E}^{2} \mathbf I-\mathbf W_{L}^{H}\left(\mathbf W_{U}+\sigma_{U}^{2} \mathbf I\right)^{-1} \mathbf W_{L}\right) \\
&-\sum_{i=1}^{M} \log _{2} \lambda_{i}\left(2 \mathbf W_{U}+\sigma_{N}^{2} \mathbf I-2 \mathbf W_{L}\left(\mathbf W_{E}+\sigma_{E}^{2} \mathbf I\right)^{-1} \mathbf W_{L}^{H}\right) \\
&-M \log _{2}\left(\sigma_{1}^{2}\right /P).
\end{aligned}
\end{aligned}
\end{equation}
\end{small}
%According to \eqref{R_BS}, $\mathbf R_{BS}$ is only related to the angle $\varphi_{B S}$, considering that the positions of IRS and BS are fixed, the elements in matrix $\mathbf R_{B S}$ are all fixed values.
Since $\operatorname{rank}(\mathbf R_{B S})=1$, $\mathbf R_{B S}$ has at most one non-zero eigenvalue, which can be denoted as $\lambda_{M}(\mathbf R_{B S})$, where $\lambda_{i}(.)$ means the $i$th eigenvalue of matrix, $\sum_{i=1}^{M} \lambda_{i}\left(\mathbf R_{B S}\right)=\lambda_{M}\left(\mathbf R_{B S}\right)=\operatorname{tr}\left(\mathbf R_{B S}\right)=M$.
%\begin{equation}
%\sum_{i=1}^{M} \lambda_{i}\left(\mathbf R_{B S}\right)=\lambda_{M}\left(\mathbf R_{B S}\right)=\operatorname{tr}\left(\mathbf R_{B S}\right)=\sum_{i=1}^{M}\left[\mathbf R_{B S}\right]_{i i}=M.
%\end{equation}
Therefore, we can have $\sum_{i=1}^{M} \lambda_{i}\left(\mathbf W_{U}\right)=M p_{U}(\mathbf v)\label{W_U}$, $\sum_{i=1}^{M} \lambda_{i}\left(\mathbf W_{E}\right)=Mp_{E}(\mathbf v)$ and $\sum_{i=1}^{M} \lambda_{i}\left(\mathbf W_{L}\right)=M p_{L}(\mathbf v)$.According to \eqref{R}, secret key rate is related to the eigenvalues of the four matrix terms composed of matrices $\mathbf W_{U}$, $\mathbf W_{E}$ and $\mathbf W_{L}$, thus we can deduce that the matrix $\mathbf W_{U}+\sigma_{U}^{2} I$ has $M-1$ identical eigenvalue $\sigma_{U}^{2}$, the $M$th eigenvalue is $M p_{U}(\mathbf v)+\sigma_{U}^{2}$. The eigenvalues of the other three matrix terms can be calculated similarly. Substituting these eigenvalues into \eqref{R}, we have $R$ as \eqref{theorem1}.
%According to \eqref{R}, secret key rate is related to the eigenvalues of the four matrices $\mathbf W_{U}+\sigma_{U}^{2} \mathbf I$, $\mathbf W_{E}+\sigma_{E}^{2} \mathbf I$, $\mathbf W_{E}+\sigma_{E}^{2} \mathbf I-\mathbf W_{L}^{H}\left(\mathbf \mathbf W_{U}+\sigma_{U}^{2} \mathbf I\right)^{-1} \mathbf W_{L}$ and $2 \mathbf W_{U}+\sigma_{N}^{2} \mathbf I-2 \mathbf W_{L}\left(\mathbf W_{E}+\sigma_{E}^{2} \mathbf I\right)^{-1} \mathbf W_{L}^{H}$, and they are all composed of matrix $\mathbf W_{U}$, $\mathbf W_{E}$ and $\mathbf W_{L}$, thus we can deduce that the matrix $\mathbf W_{U}+\sigma_{U}^{2} I$ has $M-1$ identical eigenvalue $\sigma_{U}^{2}$, the $M$th eigenvalue is $M p_{U}(\mathbf v)+\sigma_{U}^{2}$. The eigenvalues of the other three matrices can be calculated similarly. Substituting these eigenvalues into \eqref{R}, we have $R$ as \eqref{theorem1}.

	\bibliographystyle{IEEEtran}
%	\bibliography{IEEEabrv,refer}
	\bibliography{main}

\ifCLASSOPTIONcaptionsoff
  \newpage
\fi

\end{document}